\documentstyle[aps,multicol,psfig]{revtex}

\newcommand{\be}{\begin{equation}}
\newcommand{\ee}{\end{equation}}
\newcommand{\bea}{\begin{eqnarray}}
\newcommand{\eea}{\end{eqnarray}}
\newcommand \ga{\raisebox{-.5ex}{$\stackrel{>}{\sim}$}}
\newcommand \la{\raisebox{-.5ex}{$\stackrel{<}{\sim}$}}
\newcommand{\rx}{{\rm x}}
\newcommand{\ry}{{\rm y}}
\newcommand{\rz}{{\rm z}}
\begin{document}

\draft

\title{Emission times and opacities from interferometry\\ 
in non-central Relativistic Nuclear Collisions}

\author{Henning Heiselberg}

\address{NORDITA, Blegdamsvej 17, DK-2100 Copenhagen \O, Denmark}

\maketitle

\begin{abstract}
The nuclear overlap zone in non-central relativistic heavy ion
collisions is azimuthally very asymmetric. By varying the angle
between the axes of deformation and the transverse direction of the
pair momenta, the transverse HBT radii oscillate in a characteristic
way. It is shown that these oscillations allow determination of
source sizes, deformations as well as the opacity and duration of
emission of the source created in any non-central high energy nuclear
collisions. The behavior of the physical quantities with centrality
of the collisions is discussed --- in particular changes caused by a
possible phase transition to a quark-gluon plasma.
\end{abstract}


\begin{multicols}{2}

Understanding particle emission in relativistic nuclear collision is
crucial in order to determine whether a phase transition to
quark-gluon plasma has occurred. Particle interferometry was invented
by Hanbury-Brown \& Twiss (HBT) for stellar size determination
\cite{HBT} and is now employed in nuclear collisions
\cite{GKW,Pratt,Csorgo,Heinz,HH,NA49,NA44}. It is a very powerful
method to determine the 3-dimensional source sizes, life-times,
duration of emission, flow, etc. of pions, kaons, etc. at
freeze-out. Since the number of pairs grow with the multiplicity per
event squared the HBT method will become even better at RHIC and LHC
colliders where the multiplicity will be even higher. In this paper an
extension of the HBT method is presented for asymmetric and opaque sources
created in non-central collisions. It is shown that combining HBT with
a determination of the reaction plane can be exploited to find not
only the size and deformation of the source but also its opacity as
well as the duration of emission separately.

The reaction plane, which breaks azimuthal symmetry, has been
successfully determined in non-central heavy ion collisions from
intermediate \cite{Bevalac} up to relativistic energies
\cite{AGS,NA49v2}.  The particle spectra are expanded in harmonics of
the azimuthal angle $\phi$ event-by-event
\cite{Bevalac,AGS,NA49v2,Ollitrault,Voloshin,Wiedemann}
\bea
  E\frac{dN}{d^3p} &=& \frac{dN}{p_tdp_tdyd\phi} 
  = \int d^4x S(x,p)  \nonumber\\ 
    &=& \frac{dN}{2\pi p_tdp_tdy} 
   [1+2v_1\cos(\phi-\phi_R)  \nonumber\\ 
  && +2v_2\cos2(\phi-\phi_R)\;+... ]  \,, \label{v}
\eea
where $\phi_R$ is the azimuthal angle of the reaction plane. Assuming
that experimental uncertainties in event plane reconstruction can be corrected
for, 
each event can be rotated such that $\phi_R=0$. The asymmetry decreases with
centrality (see Fig. 1) and vanish for the very central collisions, which
are cylindrical symmetric. 
The standard
geometry has the $z$-direction along the longitudinal or beam axis and
the $\rx$-direction along the impact parameter ${\bf b}$. 
Thus $(\rx,\rz)$ constitutes the {\it reaction plane} and $(\rx,\ry)$
the transverse direction with $\ry$ perpendicular to the reaction
plane. In HBT analyses one often parametrize
the transverse particle distribution as a gaussian function of
transverse radius.  It is therefore convenient for asymmetric sources
to employ gaussian parametrizations in both transverse directions
\bea
  S_\perp(\rx,\ry)  \sim 
  \exp\left( - \frac{\rx^2}{2R_x^2} -  \frac{\ry^2}{2R_y^2}\right)
   \,. \label{gaussian}
\eea
Here $R_x$ and $R_y$ are the gaussian transverse sizes of the
collision zones at freeze-out. This elliptic source neglects directed flow
and we shall therefore be restricted to regions around cms midrapidity where
$v_1$ vanish \cite{AGS,NA49v2}.

The azimuthal asymmetry or ``deformation'' of the source 
can be defined as the relative difference between the gaussian radii squared
\bea
  \delta \equiv \frac{R_y^2-R_x^2}{R_y^2+R_x^2} \,. \label{delta}
\eea
A simple estimate can be obtained from the full transverse extent of the
initial overlap of two nuclei with radius $R_A$ colliding with impact parameter
$b$ (see Fig. 1). They are simply: 
$R_x=R_A-b/2$ and $R_y=\sqrt{R_A^2-b^2/4}$, 
and the corresponding deformation is
\bea
  \delta = \frac{b}{2R_A} \,.
\eea
The rms radii of the nuclear overlap zone weighted with longitudinal
thicknesses results in a deformation that is only slightly smaller
at semicentral collisions. However, as the source expands the
deformation $\delta\simeq(R_y-R_x)/(R_y+R_x)$
decreases for two reasons. 
Firstly, the expansion increase $(R_x+R_y)$ and secondly, $(R_y-R_x)$
decrease because the average velocities are larger in the $\rx$- than
$\ry$-direction.  The latter is a consequence of the experimentally
measured positive elliptic flow ($v_2>0$ in Eq. (\ref{v})) in
relativistic nuclear collisions where shadowing is minor
\cite{AGS,NA49v2}.
Measuring the decrease of the deformation with centrality will reveal 
important information on the expansion up to freeze-out. 
For very peripheral collisions, where only a single nucleon-nucleon
collision occurs, the source must be azimuthally symmetric, i.e., the
deformation must vanish and therefore Eq. (4), which assumes continuous
densities, breaks down.

The standard HBT method for calculating the Bose-Einstein correlation function
from the interference of two identical particles is now briefly discussed.
For a source of size $R$ we consider two particles emitted a distance
$\sim R$ apart with relative momentum ${\bf q}=({\bf p}_1-{\bf p}_2)$
and average momentum, ${\bf P}=({\bf p}_1+{\bf p}_2)/2$. Typical heavy
ion sources in nuclear collisions are of size $R\sim5$ fm, so that
interference occurs predominantly when 
$q\raisebox{-.5ex}{$\stackrel{<}{\sim}$}\hbar/R\sim 40$
MeV/c. Since typical particle momenta are $p_i\ga P\sim 300$ MeV/c,
the interfering particles travel almost parallel, i.e.,
$p_1\simeq p_2\simeq P\gg q$.  The correlation function due to
Bose-Einstein interference of identical spin zero bosons 
$\pi^\pm\pi^\pm$, $K^\pm K^\pm$, etc.)
from an incoherent source is (see, e.g., \cite{Heinz})
\begin{equation}
  C_2({\bf q},{\bf P})=1\; +\; \left|\frac{\int d^4x\;S(x,{\bf P})\;e^{iqx}}
   {\int d^4x\;S(x,{\bf P})}\right|^2 \,, \label{C}
\end{equation}
where $S(x,{\bf P})$ is the source distribution 
function describing the phase space density of the emitting source.

Experimentally the correlation functions are often
parametrized by the gaussian form
\bea
  C_2(q_s,q_o,q_l)&=&1+\lambda\exp[
           - q_s^2R_s^2-q_o^2R_o^2-q_l^2R_l^2  \nonumber\\
  &&  -2q_oq_sR_{os}^2 
      -2q_oq_lR_{ol}^2 -2q_lq_sR_{sl}^2 ]\;.   \label{Cexp}
\eea
Here, ${\bf q}={\bf p}_1-{\bf p}_2=(q_s,q_o,q_l)$ is the relative
momentum between the two particles and $R_i,i=s,o,l,os,ol,sl$ the
corresponding sideward, outward, longitudinal, out-sideward,
out-long and sideward-longitudinal HBT radii respectively.  
We have suppressed the ${\bf P}$
dependence.  We will employ the standard geometry, where the {\it
longitudinal} direction is along the beam axis, the {\it outward}
direction is along ${\bf P}_\perp\simeq {\bf p}_{\perp,i}$, 
and the {\it sideward} axis is
perpendicular to these. Usually, each pair of particles is lorentz
boosted longitudinal to the system where their rapidity vanishes,
$y=0$. Their average momentum ${\bf P}$ is then perpendicular to the
beam axis and is chosen as the outward direction. In this system the
pair velocity \mbox{\boldmath $\beta_P$}=${\bf P}/E_P$ points in the
outward direction with $\beta_o=p_\perp/m_\perp$, where
$m_\perp=\sqrt{m^2+p_\perp^2}$ is the transverse mass, and both the
$R_{ol}$ and $R_{sl}$ vanish at midrapidity (see Ref.
\cite{Heinz} for further analyses). 
Also $R_{os}$ vanishes for a cylindrically symmetric source or if
the azimuthal angle of the reaction plane is not determined and
therefore averaged over --- as has been the case experimentally so far.
The reduction factor $\lambda$ in Eq. (\ref{Cexp})
may be due to long lived resonances \cite{Csorgo,HH},
coherence effects, incorrect Coulomb corrections or other effects. It
is $\lambda\sim 0.5$ for pions and $\lambda\sim 0.9$ for kaons.

 The Bose-Einstein correlation function can now be calculated for a
deformed source. 
Let us first investigate {\it transparent} sources and, as in \cite{Wiedemann},
parametrize the transverse and temporal extent by gaussians  
\bea
   S(x,P) \sim S_\perp(\rx,\ry) \exp[
         -\frac{(\tau-\tau_f)^2}{2\delta\tau^2}] \, e^{p\cdot u/T} 
   \,, \label{S}
\eea
with longitudinal Bjorken flow,
$u=(\cosh\eta,0,0,\sinh\eta)$. Effects of 
transverse flow will be discussed below.
The transverse radii $R_x,R_y$ are the gaussian radii at freeze-out, 
$\tau_f$ is the freeze-out time and $\delta\tau$ the duration of emission.

In order to calculate the correlation function of (\ref{C}) the
gaussian approximation is employed (see, e.g., \cite{Csorgo})
which results in a correlation function on the form as in
Eq. (\ref{Cexp}). Inserting the source (\ref{S}) in Eq. (\ref{C}) and
Fourier transforming we obtain the correlation function.  Comparing to
the experimental parametrization of Eq.(\ref{Cexp}), 
one calculates the HBT radii. 
For transparent sources the azimuthal dependence of
the HBT radii has been calculated in detail by Wiedemann
\cite{Wiedemann}. In the longitudinal center-of-mass system
of the pair $(y=0)$, the HBT radii are around midrapidities
\bea
   R_s^2 &=& R^2\left[1 + \delta\cos(2\phi)\right] \,, \label{Rst} \\
   R_o^2 &=& R^2\left[1 - \delta\cos(2\phi)\right] 
             \,+ \beta_o^2\delta\tau^2  \,,   \label{Rot} \\
   R_{os}^2 &=& R^2\delta \sin(2\phi) \,, \label{Rost}\\
   R_l^2 &=& \frac{T}{m_\perp} \tau_f^2 \,, \label{Rl}
\eea
where $R^2=(R_x^2+R_y^2)/2$ is the average of the source radii
squared.  As in asymmetric flow, Eq. (\ref{v}), 
$\phi$ is the azimuthal angle between
the transverse momentum $p_\perp$ and the reaction plane. It is
therefore the angle by which the $R_{o,s}$ axes are rotated with
respect to the $(\rx,\ry)$ reaction plane (see Fig. 1). Near target and
projectile rapidities the directed flow is appreciable and leads to
$\cos\phi$ terms in Eqs. (\ref{Rst}-\ref{Rot}) \cite{Ollitrault,Wiedemann}.
The out- and sideward HBT radii show a
characteristic modulation as function of azimuthal angle with
amplitude of same magnitude but opposite sign. 
Measuring the amplitude modulation of $R_{s,o,os}$ determines five quantities
and thus overdetermines the three source parameters which are 
the source size  $R$, deformation $\delta$ and 
and duration of emission $\delta\tau$.

Next we consider {\it opaque} sources. In relativistic heavy ion
collisions source sizes and densities are large and one would expect
rescatterings. As a result particles are predominantly emitted near
the surface and arrive from the (front) side of the source facing
towards the detector.  In \cite{Opaque} it was found that for opaque
sources, where mean free paths are smaller than source sizes,
$\lambda_{mfp}\la R$, the sideward HBT radius increase whereas the
outward is significantly reduced.  The simple geometrical cause is
that the emission region in the outward direction is the surface
region which considerably narrower than the whole source.  As in
\cite{Opaque,Tomasik} Glauber absorption is introduced by adding an
absorption factor $\exp(-\int_x \sigma\rho(x')dx')$ where $\sigma$ is
the interaction cross section, $\rho$ the density of scatterers and
the integral runs along the particle trajectory from source point $x$
to the detector. Defining the mean free path as
$\lambda_{mfp}=(\sigma\rho(0))^{-1}$, where $\rho(0)$ is the central
density, the source is opaque when $\lambda_{mfp}\ll R$ and
transparent when $\lambda_{mfp}\gg R$.  Calculating the correlation
function for an opaque source from Eq. (\ref{C}) and comparing to the
definition of the HBT radii in Eq. (\ref{Cexp}), one generally obtains
by expanding for small deformations,
\bea
   R_s^2 &=& g_sR^2\left[1 + \delta\cos(2\phi)\right] \,, \label{Rso} \\
   R_o^2 &=& g_oR^2\left[1 - \delta\cos(2\phi)\right] 
             \,+ \beta_o^2\delta\tau^2  \,,           \label{Roo} \\
   R_{os}^2 &=& g_{os} R^2\delta \sin(2\phi) \,. \label{Roso}
\eea
Here $g_{o,s,os}$ are model
dependent factors that are functions of opacity but
independent of the deformation. 
For a gaussian source ($\rho\propto S_\perp$), 
which is moderately opaque $(\lambda_{mfp}/R=1$),
a numerical calculation gives 
$g_s\simeq 1.4$ and $g_o\simeq 0.9$ (see also \cite{Tomasik}).
For a disk source with transverse radius twice the gaussian radius $2R$,
that emits like a black body (i.e., $\lambda_{mfp}\ll R$),
one finds $g_s=4/3$ and $g_o=4(\frac{2}{3}-(\frac{\pi}{4})^2)\simeq 0.2$
\cite{Opaque}. In all cases $g_{os}\simeq g_s$. Generally, 
$(g_s-g_o)\ge 0$ and the difference increases with opacity. 
Only for a completely transparent sources is $g_s=g_o$.
In Fig. 2 the HBT radii of
Eqs. (\ref{Rst}-\ref{Roso}) are shown for a near-central collisions
($\delta=0.2$) with a moderate duration of emission
($\beta_o^2\delta\tau^2/g_sR^2=1/4$) for various opacities
$\lambda_{mfp}/R=0.1,0.5,1.0,2.0,\infty$. 
As the opacity increases, $g_o/g_s$ decreases and therefore also
the outward HBT radius and its amplitude.

Comparing the HBT radii from an opaque source
Eqs.(\ref{Rso}-\ref{Roso}) with those of a transparent source
Eqs.(\ref{Rst}-\ref{Rost}), one notices that {\it the amplitudes in
$R_s$ and $R_o$ differ} by the amount $(g_s-g_o)$.
The modulation of the HBT radii with $\phi$
provides five measurable quantities which over-determinates
the four physical quantities describing the source: its size $R$, deformation
$\delta$, opacity $(g_o-g_s)$ and duration of emission $\delta\tau$, at
each impact parameter.  The azimuthal dependence of the HBT radii thus
offers an unique way to determine the opacity of the source as well as
the duration of emission separately.  

Experimentally, HBT analyses have not been combined
with determination of the reaction plane yet. Consequently, the azimuthal
angle $\phi$ is averaged and information on three of five measurable
quantities in Eqs. (\ref{Rso}-\ref{Roso}) is lost. From the angular averaged
difference between the out- and sideward HBT radii
\bea
 \langle R_o^2-R_s^2\rangle_\phi = \beta_o^2\delta\tau^2 - (g_s-g_o)R^2 
 \,, \label{Ro-Rs}
\eea
one can only determine the sum of the positive 
duration of emission and negative opacity effect.
Experimentally the difference is small; NA49 \cite{NA49} and
NA44 \cite{NA44} data even differ
on the sign. Detailed analyses of the $p_\perp$ dependence of the HBT radii
from NA49 data within opaque sources
\cite{Tomasik} indicate that the sources are transparent or at most moderately
opaque. However, the NA44 data for which $R_o\la R_s$ necessarily requires
an opaque source as seen from Eq. (\ref{Ro-Rs}).
Furthermore, the $p_\perp$ dependence of the transverse 
HBT radii change if the source sizes, opacities, and duration of emission
also are $p_\perp$-dependent.

Transverse flow may affect the out- and sideward HBT radii as opacity,
i.e., the factors $g_{o,s}$ may depend on both.  For transparent
sources transverse flow has been studied in
\cite{Wiedemann,Csorgo,Tomasik} through the flow
$u=(\gamma\cosh(\eta),\gamma\sinh(\eta),{\bf u}_\perp)$, where
$\gamma=\sqrt{1+u_\perp^2}$, assuming that the transverse flow scales
with transverse distance, ${\bf u}_\perp\simeq u_0 {\bf
r}_\perp/R$. Otherwise the same transparent gaussian source as in
Eq. (\ref{S}) was employed. To lowest order in the transverse flow
both transverse HBT radii decrease by the same factor
$\sim(1+u_0^2m_\perp/T)$ to leading order in $u_0$.  This transverse
flow correction is independent of the source size and therefore also
the deformation. Consequently, spatial deformations reduce the
amplitudes by the same amount in this model for both transparent (see
\cite{Wiedemann}) and opaque sources.  There are, however, box shape
models where the transverse flow reduce $R_o$ more than $R_s$
\cite{Tomasik}. A similar flow effect is found in hydrodynamic models
also \cite{Rischke} although it is considerably less than the opacity
effect, i.e., when the Cooper-Frye freeze-out condition is modified by
removing the back side of the source.  Preferrably, the $p_\perp$
dependence of the HBT radii should be measured as well as the
transverse flow from apparent temperatures in order to determine the
magnitude of the flow and opacity separately.  The transverse flow
might also be azimuthally dependent, i.e., $g_{s,o}$ depend on $\phi$
and thereby change the amplitude. This effect can be estimated from
the measured elliptic flow and we find that for semicentral $Pb+Pb$
collisions at SPS energies $\langle u_x^2\rangle-\langle
u_y^2\rangle\sim v_2$, which is only a few percent. Therefore, the
azimuthally dependence of the flow and its effect on the amplitude in
the HBT radii is also of that order only which is much less than
$\delta\simeq 0.5$ for semi-central collisions. The conclusion is that
besides opacity also transverse flow may affect the factors $g_{o,s}$
- but independently of azimuthal angle. Therefore
Eqs. (\ref{Rso}-\ref{Roso}) are still valid and can be used to extract
the duration of emission unambiguously.

Measuring the centrality or impact parameter dependence of the source
sizes, deformation, opacity, emission times and duration of emission
is very important for determining how the source change with initial
energy density.  If no phase transition takes place one would expect
that source sizes and emission times increase gradually with
centrality whereas the deformation decrease approximately as
(\ref{delta}).  In peripheral collisions source sizes and densities
are small and few rescatterings occur. Therefore, the source is
transparent and the HBT radii are given by Eqs. (\ref{Rst}-\ref{Rl}).
For near central collisions sources sizes and densities are higher
which leads to more rescatterings. Thus the source is more opaque and
the amplitudes should differ. It would be interesting to observe this
gradual change in amplitude with centrality. At the same time it would
provide a direct experimental determination whether the source is
transparent or opaque as well as extracting the magnitudes of the
opacity and duration of emission.

If a phase transition occurs at some centrality, where energy
densities exceed the critical value, one may observe sudden changes in
these quantities. The emission time and duration of emission has in
hydrodynamic calculations \cite{Rischke} been predicted to increase
drastically leading to very large $R_l$ and $R_o$.  A long lived mixed
phase would also emit particles as a black body and thus the opacity
should also become large. If droplet formation occurs leading to
rapidity fluctuations, one may be able to trigger on such fluctuations
and find smaller longitudinal and sideward HBT radii \cite{HJ}.
Furthermore, if an interesting change in these quantities should occur
at some centrality, it would also be most interesting to look for
simultaneous $J/\Psi$ suppression, strangeness enhancement, decrease
in directed, elliptic or transverse flow \cite{HL}, 
or other signals from forming a quark-gluon plasma.

In summary, the importance of measuring the reaction plane and HBT
radii simultaneously has been stressed. The modulation of the HBT
radii with azimuthal angle between the reaction plane and particle
transverse momenta can be exploited to obtain source sizes,
deformations, life-times, duration of emission and opacities
separately. Tracking these physical quantities with centrality will
provide detailed information about the source created in relativistic
nuclear collisions and may reveal the phase transition to quark-gluon
plasma.


\end{multicols}


\begin{figure}
\centerline{\psfig{figure=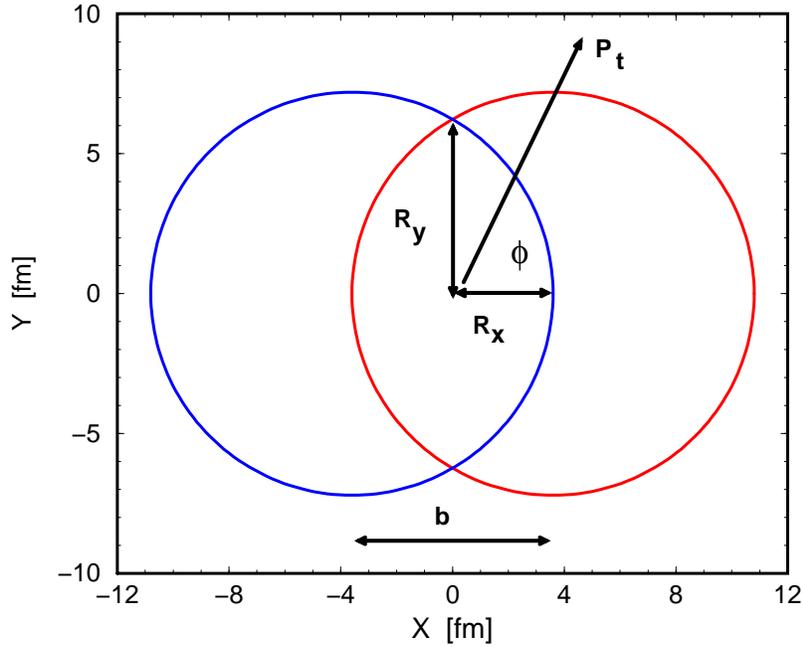,width=12cm,height=10cm,angle=-90}}
\vspace{0.5cm}
\caption{Reaction plane of semi-central $Pb+Pb$ collision for impact parameter
$b=R_{Pb}\simeq7$fm. The overlap
zone is deformed with $R_x\le R_y$. The reaction plane $(x,z)$ is
rotated by the angle $\phi$ with respect to the transverse particle momentum
$p_\perp$ which defines the outward direction in HBT analyses.
\label{geofig}  }
\end{figure}

\vspace{-1.8cm}
\begin{figure}
\centerline{\psfig{figure=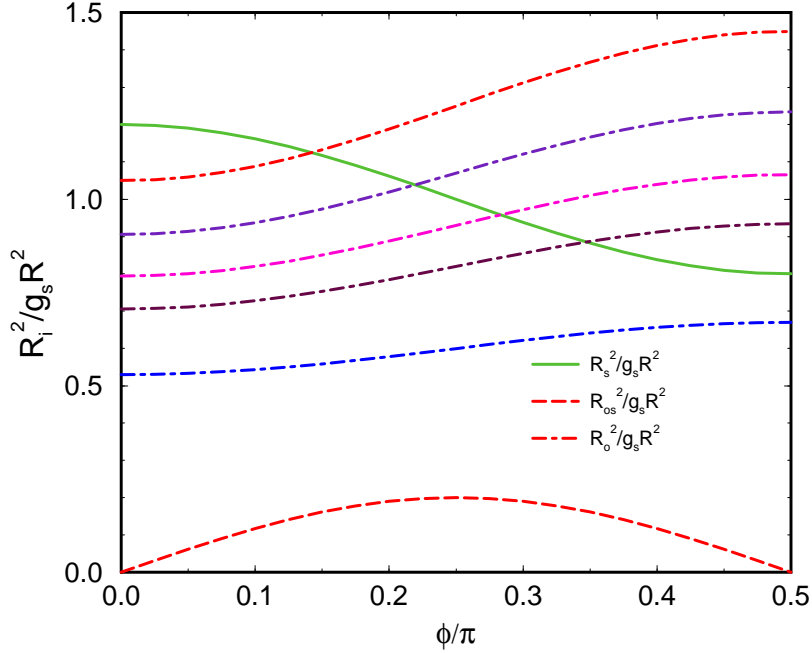,width=12cm,height=10cm,angle=-90}}
\vspace{0.5cm}
\caption{HBT radii vs. angle between reaction plane and transverse
particle momenta. The HBT radii are normalized 
to the angle averaged sideward HBT radius
squared, $g_sR^2$, for slightly deformed $\delta=0.2$ source with duration
of emission $\beta_o^2\delta\tau^2/g_sR^2=1/4$ and with various opacities. 
The sideward HBT radius (full curve) is then the
same for both transparent (Eq. (\ref{Rst})) and opaque (Eq. (\ref{Rso}))
sources and likewise for the out-side HBT radii (Eqs. (\ref{Rost}) and 
(\ref{Roso}), dashed curve). 
The outward HBT radii are shown with  chain-dashed curves
for a gaussian source (Eq. (\ref{Roo}) and (\ref{Rot}) for various
opacities (from below and up): $\lambda_{mfp}/R=0.1,0.5,1.0,2.0,\infty$.
\label{R2fig}  }
\end{figure}

\end{document}